\documentclass[manuscript]{aastex}
\shorttitle{CME flare relation}
\shortauthors{Bein et al.}

\usepackage{natbib}
\usepackage{graphicx}
\usepackage{amsmath}
\usepackage{rotating}

\begin{document}
\title{The height evolution of the `true' CME mass derived from STEREO COR1 and COR2 observations}

\author{B.M. Bein\altaffilmark{1},  M. Temmer\altaffilmark{1} }
\affil{$^1$Kanzelh{\"o}he Observatory-IGAM, Institute of Physics, University of Graz, Universit{\"a}tsplatz 5, 8010 Graz, Austria}

\author{A. Vourlidas\altaffilmark{2}}
\affil{$^2$Space Science Division, Naval Research Laboratory, Washington, DC, USA}

\author{A.M. Veronig\altaffilmark{1}}
\affil{$^1$Kanzelh{\"o}he Observatory-IGAM, Institute of Physics, University of Graz, Universit{\"a}tsplatz 5, 8010 Graz, Austria}

\and

\author{D. Utz\altaffilmark{3}\altaffilmark{,1}}
\affil{$^1$Kanzelh{\"o}he Observatory-IGAM, Institute of Physics, University of Graz, Universit{\"a}tsplatz 5, 8010 Graz, Austria}
\affil{$^3$IAA - Instituto de Astrof\'{i}sica de Andaluc\'{i}a, CSIC, Glorieta de la Astronom\'{i}a, s/n, 18080 Granada, Spain}

\date{Received/Accepted}

\begin{abstract}
Using combined STEREO-A and STEREO-B EUVI, COR1 and COR2 data, we derive deprojected CME kinematics and CME `true' mass evolutions for a sample of 25 events that occurred during December 2007 to April 2011. We develop a fitting function to describe the CME mass evolution with height. The function considers both the effect of the coronagraph occulter, at the beginning of the CME evolution, and an actual mass increase. The latter becomes important at about 10 to 15 R$_{\odot}$ and is assumed to mostly contribute up to 20~R$_{\odot}$. The mass increase ranges from 2 to 6\% per R$_{\odot}$ and, is positively correlated to the total CME mass. Due to the combination of COR1 and COR2 mass measurements, we are able to estimate the `true' mass value for very low coronal heights ($<3$~R$_{\odot}$). Based on the deprojected CME kinematics and initial ejected masses, we derive the kinetic energies and propelling forces acting on the CME in the low corona ($<3$~R$_{\odot}$). The derived CME kinetic energies range between $1.0-66\cdot10^{23}$ J, and the forces range between $2.2-510\cdot10^{14}$~N.

\end{abstract}


\maketitle

\section{Introduction}

The Sun's atmosphere is losing mass continuously via the solar wind outflow and sporadically via coronal mass ejections (CMEs). In this paper we focus on the mass ejected via CMEs. CME velocities derived from coronagraphic observations range from $\sim$100 to 3000~km~s$^{-1}$ \citep[e.g.][]{yashiro2004, gopalswamy2009}. It was found that the CME peak acceleration may occur very low in the corona \citep[$<$ 1.5 R$_{\odot}$ in heliocentric distances; e.g.][]{maricic2007, bein2011}. It is therefore necessary to consider also non-coronagraphic observations for the study of the CME impulsive phase. We combine EUV images with coronagraphic observations to obtain detailed CME kinematics from its initiation close to the solar surface up to about 15 R$_{\odot}$, including all three phases of CME evolution (initiation, impulsive acceleration, propagation phase) described in \citet{zhang2001, zhang2004}.
\\
\citet{vourlidas2010} analysed a sample of 7668 CMEs observed by the LASCO/SOHO coronagraphs during solar cycle 23 and found that the distribution of CME masses ranges from 10$^{13}$ to 10$^{16}$ g, with a few outliers up to 10$^{17}$ g and as small as $10^{11}$ g. When observing CMEs from one spacecraft, i.e.\ from only one vantage point, we observe the projection of a three dimensional structure on the plane-of-sky (POS) and the derived CME quantities, like the CME velocity, acceleration or mass are underestimated. The larger the deviation of the propagation direction of the CME from the POS, the larger the projection effects. With the advent of the STEREO mission it is now possible to derive propagation directions and deprojected CME quantities using various methods \citep[e.g.][]{mierla2010}. Triangulation methods estimate the direction of CMEs by combining height-time curves derived from two vantage points \citep[e.g.][]{dekoning2009, liewer2009, temmer2009}.
In the forward modeling method applied to stereoscopic observations, a flux rope is fitted to the white light observations, which enables us to estimate the CME propagation direction as well as its 3D geometry \citep{thernisien2009}. \citet{colaninno2009} presented a technique to estimate the 3D, i.e.\ deprojected, CME mass, by assuming that the difference in the observed mass from two vantage points is solely due to projection effects. We would like to note that although stereoscopic observations are used, we still do not know the real 3D structure of the CME body, and therefore, the `true' mass remains uncertain. Similar to the triangulation method, the masses observed from both vantage points are corrected for different propagation directions until both mass estimates yield the same result. Thus also the CME propagation direction can be derived with this method.
\\
Based on these techniques we present a study of deprojected CME quantities for a set of 25 events observed by both STEREO spacecraft. The set of CMEs was chosen according to observational criteria: CMEs must be well observed in both STEREO spacecraft with the instruments EUVI, COR1, and COR2. With this we are able to study the CMEs starting from the solar surface up to 15 $R_{\odot}$ and to derive detailed deprojected height-time, velocity-time and acceleration-time curves. We show how COR1 and COR2 measurements can be combined to track the CME mass evolution over a large height range. Due to the occulter disk of coronagraphs, measurements of the early evolution of the CME mass do not reflect the real CME mass accurately. Therefore we develop a fit function which considers both geometric and physical changes in the CME mass evolution. From this fit function we derive the mass at the earliest observational evolution of the CME as well as define a final CME mass, i.e. the CME reaches a constant mass above some height. Furthermore we present deprojected CME kinetic energies and forces in the low corona.

\section{Data}

The Sun-Earth-Connection Coronal and Heliospheric Investigation (SECCHI) package \citep{howard2008} onboard the Solar Terrestrial Relations Observatory (STEREO) mission \citep{kaiser2008} provides stereoscopic CME observations from the Sun up to 1 AU, by the combination of five different instruments: one Extreme Ultraviolet (EUV) Imager \citep[EUVI;][]{wuelser2004}, two white light coronagraphs (COR1 and COR2) and two white light heliospheric imagers \citep[HI1 and HI2;][]{eyles2009}. For the present study, COR1 and COR2 coronagraphic observations in combination with EUVI data are used to derive 3D CME kinematics and mass evolutions. \\
With a field-of-view (FoV) of 1.7 $R_{\odot}$, EUVI observes the upper chromosphere and lower corona in four different wavelength bands. For our study we use 195$~\rm\AA~$ images with a time cadence of 2.5 to 10 minutes. COR1 observes the inner corona between 1.4 to 4 $R_{\odot}$ with a time cadence of 5 to 10 minutes. The FoV of COR2 observations ranges from 2.5 to 15 $R_{\odot}$. We use total brightness images either obtained directly or derived from polarization sequences to achieve the maximum COR2 cadence of 15 mins. For 3 events in our sample we have lower time cadences of 30 to 60 minutes. The analysis comprises a set of 25 CME events which occurred between December 2007 and April 2011.

\section{Analysis}

For the sake of brevity, we demonstrate our data analysis procedures using a single example, the 2010 April 3 CME.

\subsection{Kinematics}
\begin{figure*}
	\centering
		\includegraphics[scale=1.2]{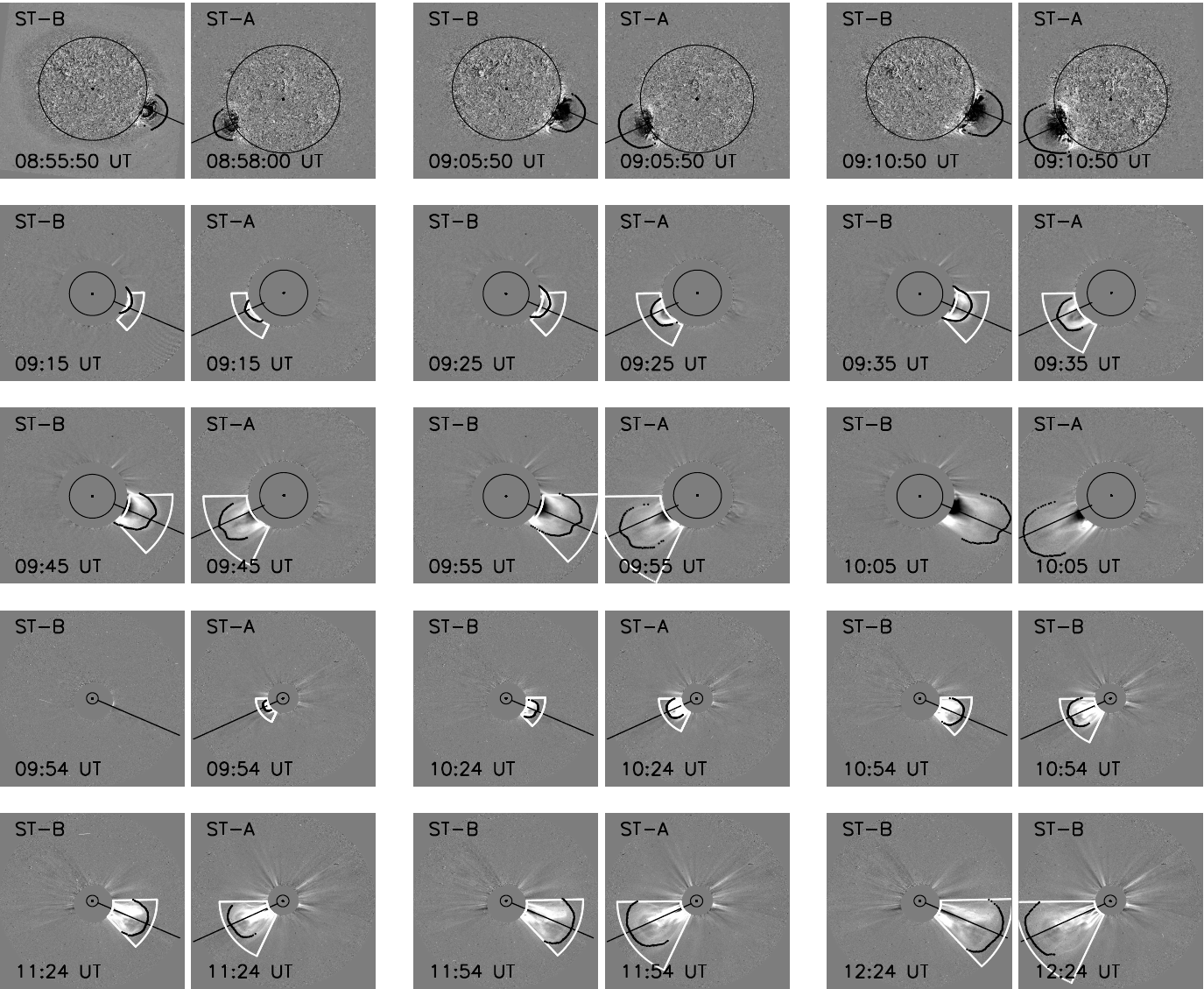}
	\caption{STEREO-A (ST-A) and STEREO-B (ST-B) observations of the CME that occurred on 2010 April 3. The top row shows running difference EUVI images, the second and third row COR1 mass images and the last two rows COR2 mass images. The temporal evolution of the event can be seen from left to right and top to bottom.
The identified CME leading edge and POS propagation direction are overplotted in black. The sectors used for the CME mass calculations are outlined in white. }
	\label{sequenz}
\end{figure*}

\begin{figure}
	\centering
		\includegraphics[scale=1.1]{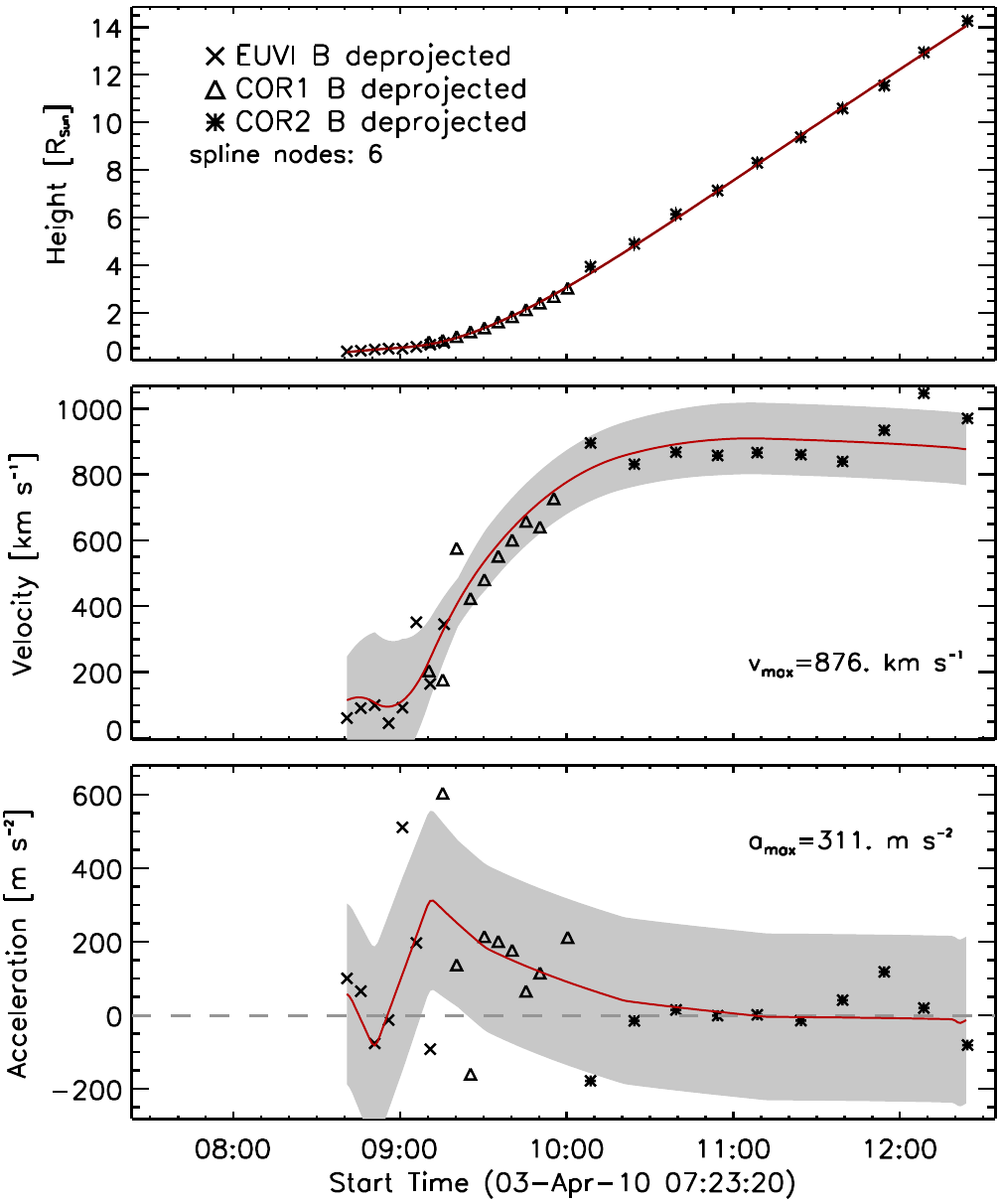}
	\caption{Deprojected kinematics of the 2010 April 3 CME. The top panel shows the height of the CME leading edge, measured from the solar surface at the location of the CME source region, against time. The plot symbols (crosses for EUVI, squares for COR1 and asterisks for COR2) represent the measurement points, the red solid line a spline fit to the data. Measurement errors (0.03 R$_{\odot}$ for EUVI, 0.125 R$_{\odot}$ for COR1 and 0.3 R$_{\odot}$ for COR2) are plotted but are smaller than the plot symbols in many cases. The first and second time derivative of the measurement points and spline fit are shown in the middle and bottom panel, representing the velocity and acceleration profile of the CME.}
	\label{kin}
\end{figure}

We measure the height-time profile of the CME leading edge in EUVI, COR1 and COR2 images to derive detailed CME kinematics starting from the solar surface up to about 15 $R_{\odot}$ using the methods described in \citet{bein2011}. Figure \ref{sequenz} shows an image sequence for the 2010 April 3 event, observed by EUVI, COR1 and COR2 from STEREO-A and STEREO-B. In each image we determine the leading edge (black solid lines in Fig.\ \ref{sequenz}) and the plane-of-sky (POS) propagation direction (black radial lines in Fig.\ \ref{sequenz}), which cross at the center of the CME front. The projected CME height is measured from STEREO-A as well as from STEREO-B observations. As reference point we identify from flaring signatures in EUVI data the location of the source region of the CME at the solar surface. From this reference point we calculate the average height for all measurement points on the CME leading edge, that are within $\pm5^\circ$ of the POS propagation direction. By applying the forward modeling of a flux rope structure to STEREO-A and STEREO-B coronagraphic observations, we obtain the propagation direction of the CME (Thernisien et al., 2009). This direction is used to correct the height-time curve in order to get deprojected kinematics.
\\
Figure \ref{kin} shows the resulting deprojected kinematics for the 2010 April 3 event. The top panel shows the height-time curve, the middle panel the CME velocity and the bottom panel the acceleration profile. Our typical measurement error is 0.03 $R_{\odot}$ for EUVI, 0.125 $R_{\odot}$ for COR1 and 0.3 $R_{\odot}$ for COR2. Our best estimate of the uncertainties due to the measurement in different wavelengths (EUV, white light) across the instruments are included in the measurement errors. From a spline fit on the derived height-time curve, we obtain smooth velocity and acceleration profiles from the first and second derivative, respectively. The maximum value in the CME acceleration profile is used as peak acceleration $a_{max}$. The peak velocity $v_{max}$ is defined as that value in the velocity profile, where the acceleration has decreased to 10\% of its maximum value. The grey shaded area in the velocity-time and acceleration-time plot represents the error range on the spline fit. For more details see \citet{bein2011}.

\subsection{Mass measurements}
Applying the technique described in \citet{vourlidas2010} on COR1 and COR2 observations, we derive the CME mass evolutions from both STEREO vantage points. Base difference images are constructed by subtracting a pre-event image, containing no CME signatures or other disturbances. Both the pre-event image and the CME image itself are corrected for instrumental effects and calibrated in units of mean solar brightness \citep[e.g.][]{poland1981}. In the ideal case this procedure results in an image, which shows a brightness excess caused by the CME only. The brightness excess is converted into electron excess with the Thomson scattering formulation assuming that all electrons are located on the spacecraft POS and a composition of 90\% H and 10\% He. Because the Thomson scattering along the line-of-sight (LOS) is the strongest for electrons located on the POS, this method leads to an underestimation of the mass of the order of a factor of 2 \citep{vourlidas2010}. Depending on the width and the propagation direction of the CME, this can affect the entire CME or parts of it.
\\
The second to fifth row in Fig.\ \ref{sequenz} show examples of mass images, in which the CME regions are outlined by using the `sector' method. We define in each image a sector, which contains the CME structure. For all observations (COR1 and COR2) from the same STEREO vantage point and relating to the same event, the width of the CME sector is kept the same. For COR1 the lower boundary of the sector is set at 0.1 R$_{\odot}$ above the occulter, for COR2 at 0.2 R$_{\odot}$ above the occulter to avoid the increased noise around the occulter disk. The upper boundary of the sector is dependent on the height of the CME leading edge. Because each pixel value of the mass image indicates the mass along the LOS at that point, we obtain the CME mass by summing up all the pixel values within the defined sector.
\\
Assuming that the same mass is observed from both spacecraft, mass measurements from STEREO-A and STEREO-B are corrected for different CME propagation directions until a best match is found \citep{colaninno2009}. This method is applied on the entire time series of images to derive a stable propagation direction. A fixed value (mean value of the propagation direction derived from the last three COR2 observations) is used to derive the 3D mass evolution for COR1 and COR2 observations.

\subsection{Error Analysis}
Several factors may affect our mass measurements. To get a lower and upper limit of the CME mass measurements, we compare for several sample events the results derived from the region of interest (ROI) method with the results derived from the `sector' method. The ROI is manually selected enclosing the boundary of the CME structure. In cases, where the boundary of the CME is blurred, we defined the ROI very tight, to get a lower limit of the CME mass measurements. The `sector' method includes all parts of the CME but sometimes also coronal disturbances (e.g.\ streamer deflections), so it serves as an upper limit.
The differences in the absolute mass estimates derived from the two methods are on average 30\%. The error due to subtracting different pre-event images, which differ in time up to 13 hours, is found to be of the order of 10-20\%.
Generally we select our pre-event images very carefully, thus the actual uncertainties are much lower. To consider the maximal possible error we apply for all events error bars of $\pm$15\% of the current mass value on our measurements. \\
\citet{vourlidas2010} discussed several other effects influencing mass measurements in LASCO images (instrumental effects, coronal background, composition of the coronal material) and showed that they are in the order of some percent, i.e.\ within our error estimate. In contrast, projection effects can lead to an underestimation of the mass of a factor of two.
\\
We minimize these projection effects with the calculation of the 3D mass evolution. However, there are still some uncertainties in the determined propagation direction. The propagation direction, which is used for calculating the 3D mass, was taken from the mass calculation method. To estimate an appropriate error, we also use propagation directions derived from other methods (triangulation, forward modeling) and compare the masses derived with the different directions. Only for five events out of our study these differences are larger than 15\% ($\sim$20\% for 2008 April 26, 2010 August 2; $\sim$30\% for 2010 September 2, 2011 January 31, 2011 March 21).
Compared to the rest of the sample under study, those events were found to be either very wide and faint (2010 September 2, 2011 January 31) or that their propagation directions deviate more than $50^\circ$ from the spacecraft POS (2008 April 26, 2010 August 27, 2011 March 21).

\section{Results}

\begin{figure}
	\centering
		\includegraphics[scale=1.2]{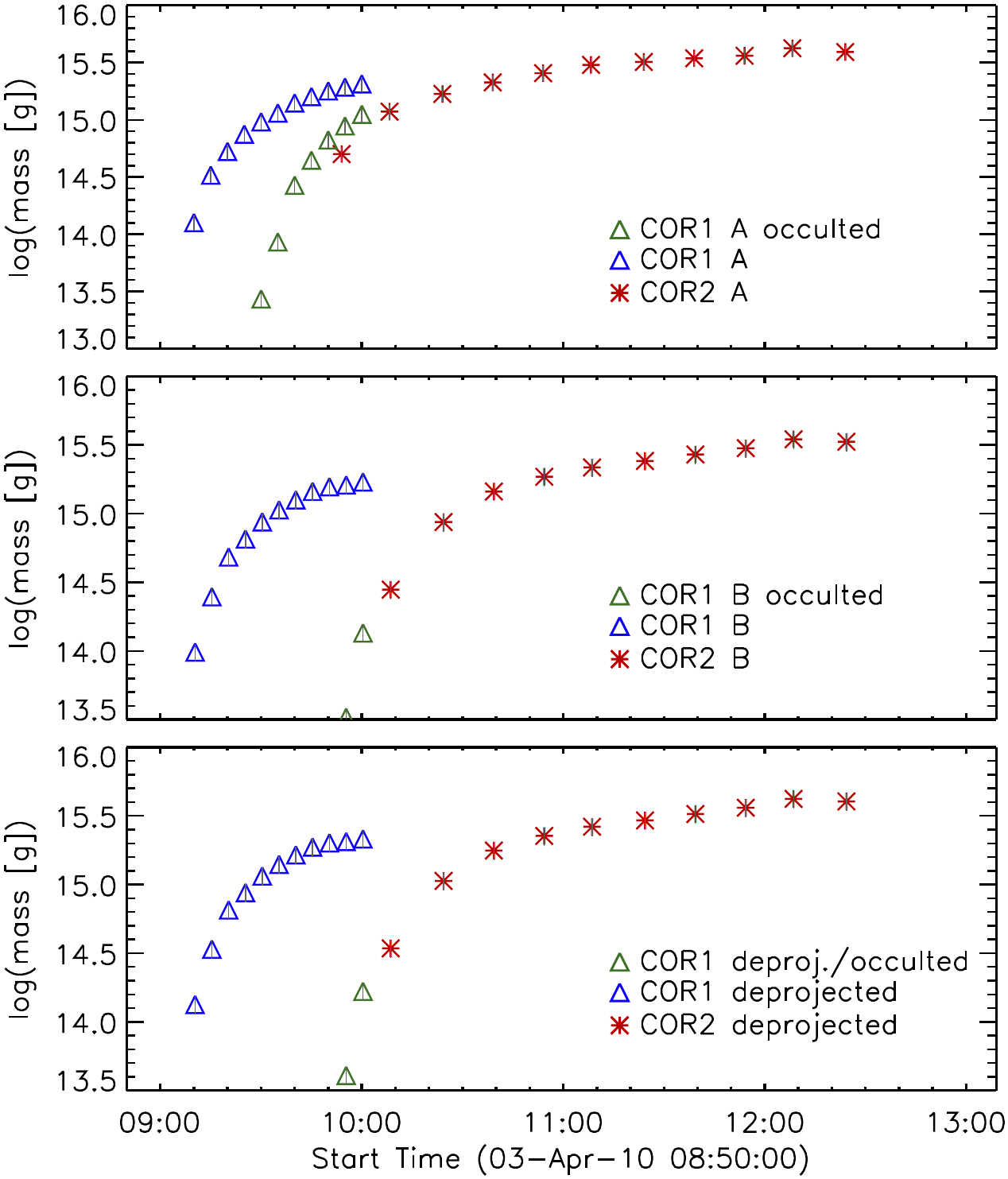}
	\caption{CME mass evolution with time for the 2010 April 3 event. Blue triangles represent COR1 measurements, green triangles mass measurements derived from COR1 observations with applying a COR2 occulter. The red asterisks represent COR2 mass observations. The top panel shows measurements derived from STEREO-A observations, the middle panel from STEREO-B observations and at the bottom panel the 3D  CME mass evolution is shown.}
	\label{masstime}
\end{figure}

\begin{figure}
	\centering
		\includegraphics[scale=0.7]{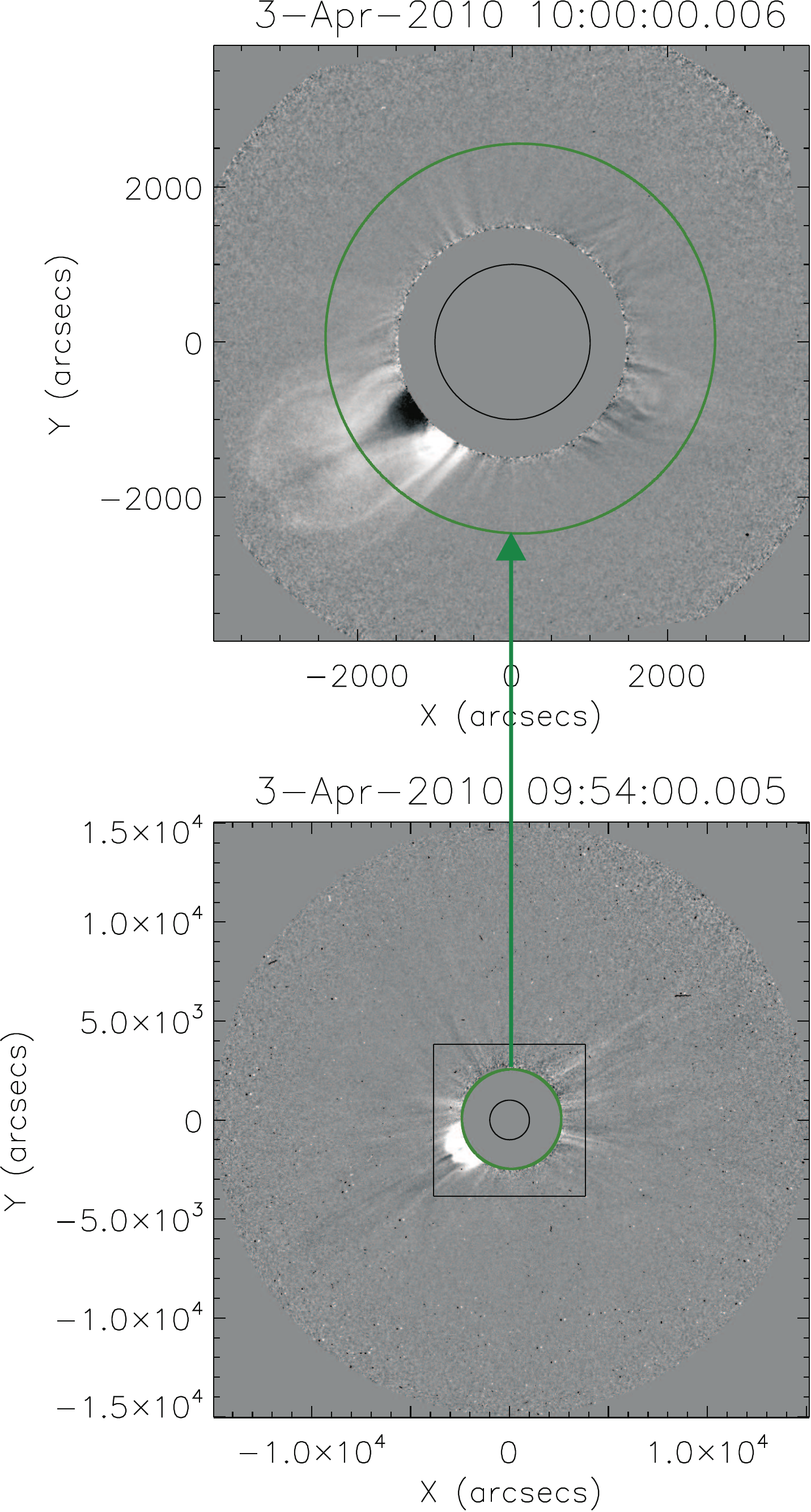}
	\caption{Mass images from the 2010 April 3 event derived from COR1 (top) and COR2 (bottom) images recorded close in time. The size of the solar disk is represented by a black circle, the size of the COR2 occulter as the green circle. The black square in the COR2 image shows the FOV of the COR1 image plotted in the upper panel.}
	\label{occulter}
\end{figure}

\begin{figure}
	\centering
		\includegraphics[scale=0.8]{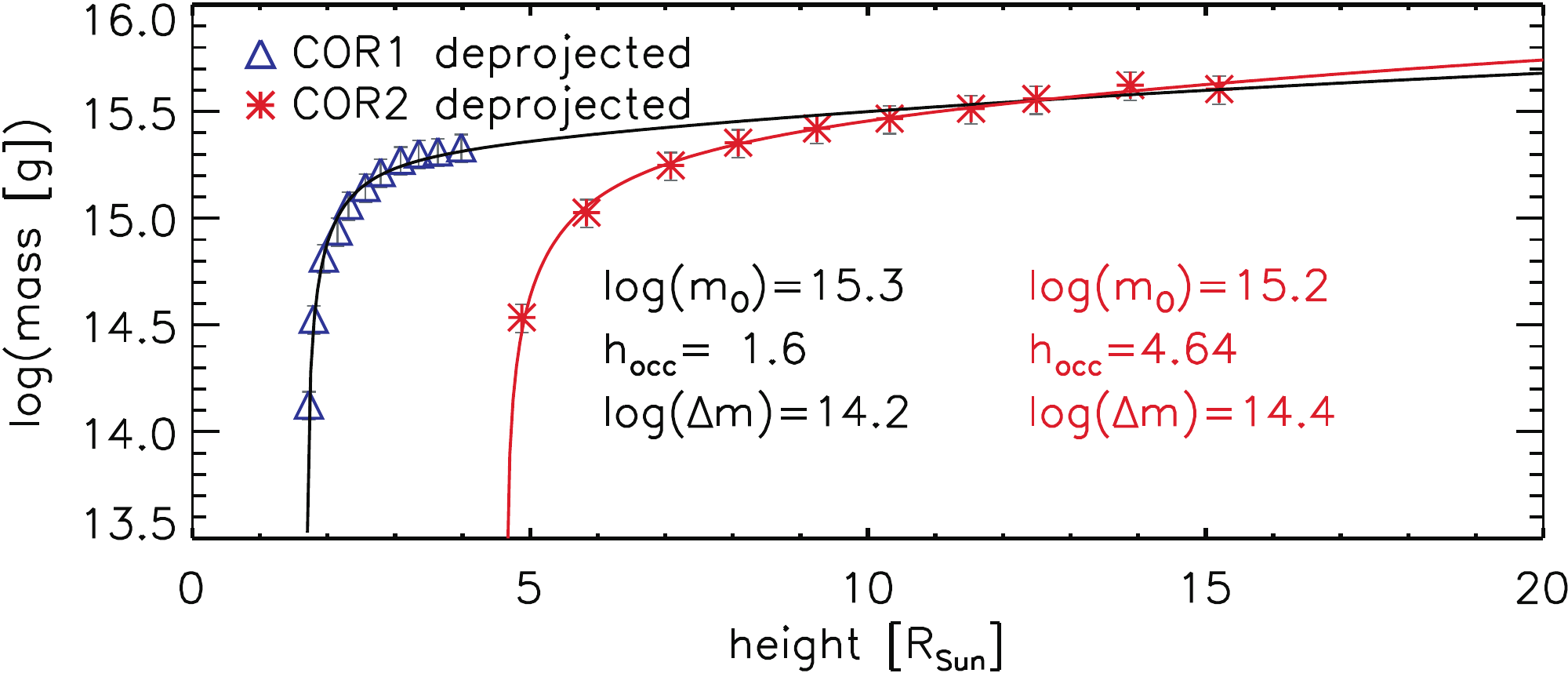}
	\caption{3D CME mass evolution against deprojected height for the CME observed on 2010 April 3, derived from COR1 observations (blue triangles) and COR2 observations (red asterisks). A fit to the data points together with the fit parameters are plotted in red for the COR2 observations and in black for the combination of COR1 and COR2 measurements. For the latter fit, COR1 measurements and all COR2 measurements, which are at least 10\% larger than the last COR1 mass measurement, are used.}
	\label{massheight}
\end{figure}

\begin{figure}
	\centering
		\includegraphics[scale=0.4]{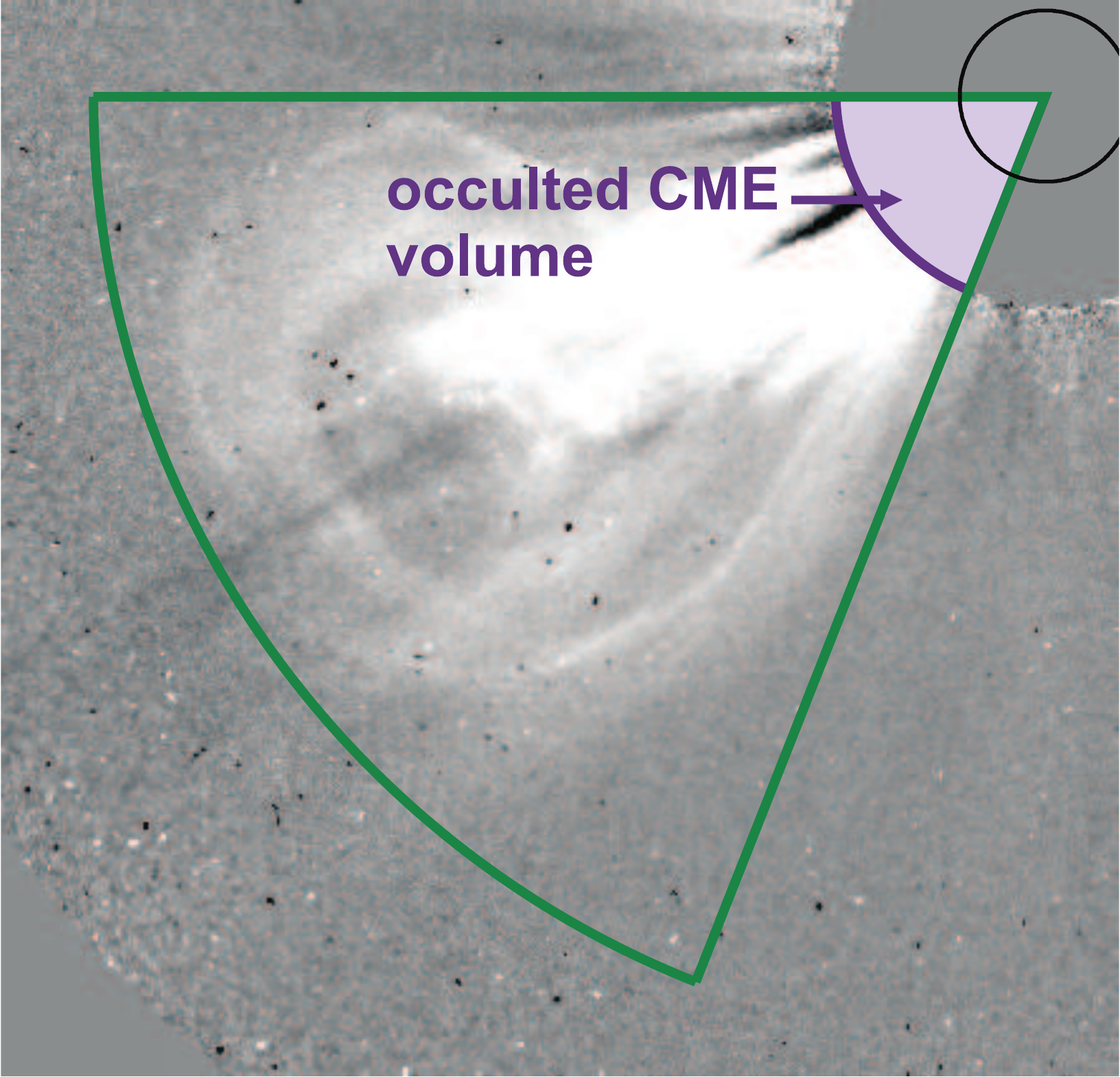}\\~\\~\\
		\includegraphics[scale=0.4]{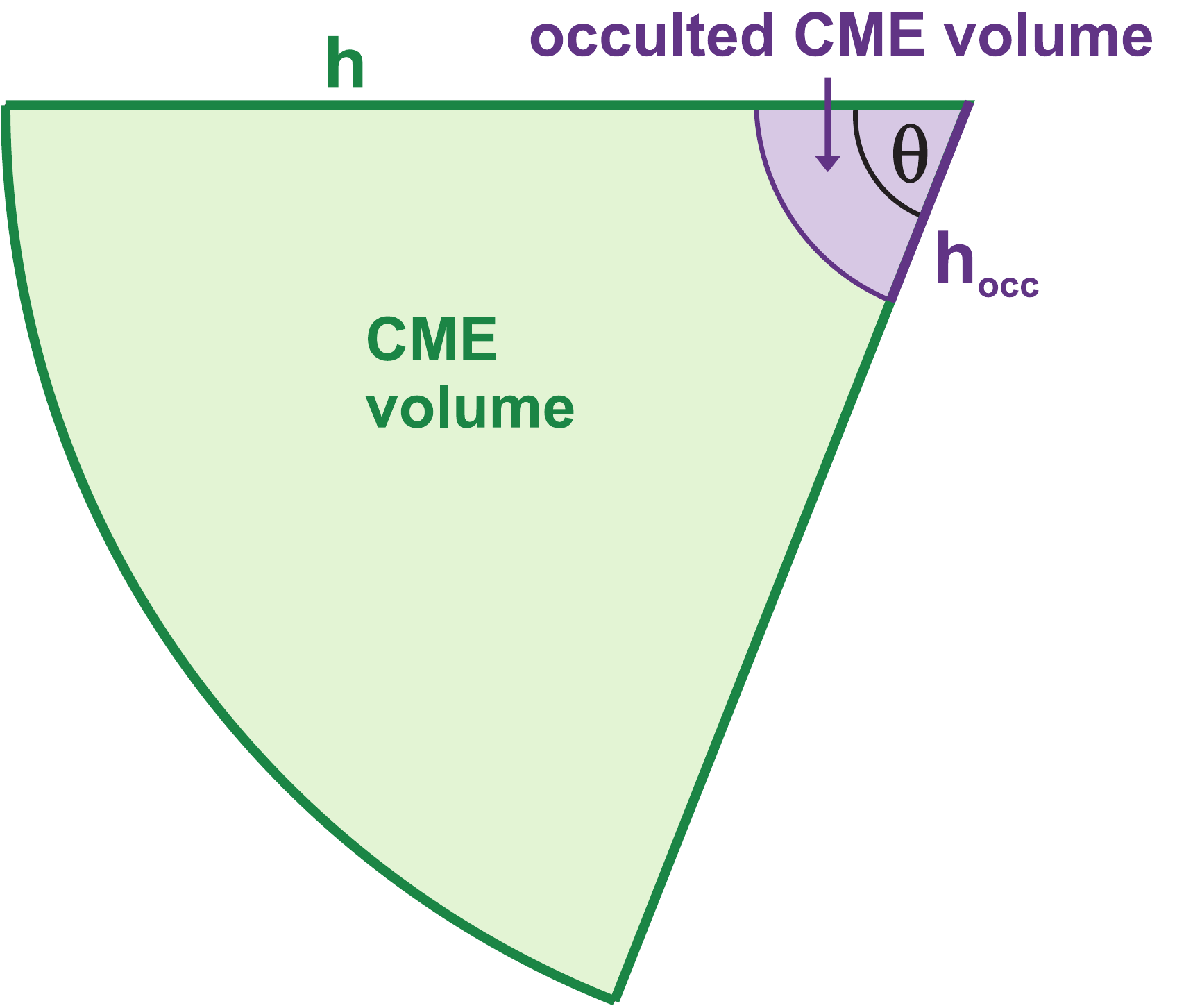}
	\caption{Top: CME image with overlaid boundary of a spherical sector (green). Bottom: spherical sector, representing the CME volume (green) with height $h$ and spherical sector representing the occulted material (purple) with height $h_{\rm occ}$.}
	\label{model1}
\end{figure}

\begin{figure}
	\centering
		\includegraphics[scale=0.8]{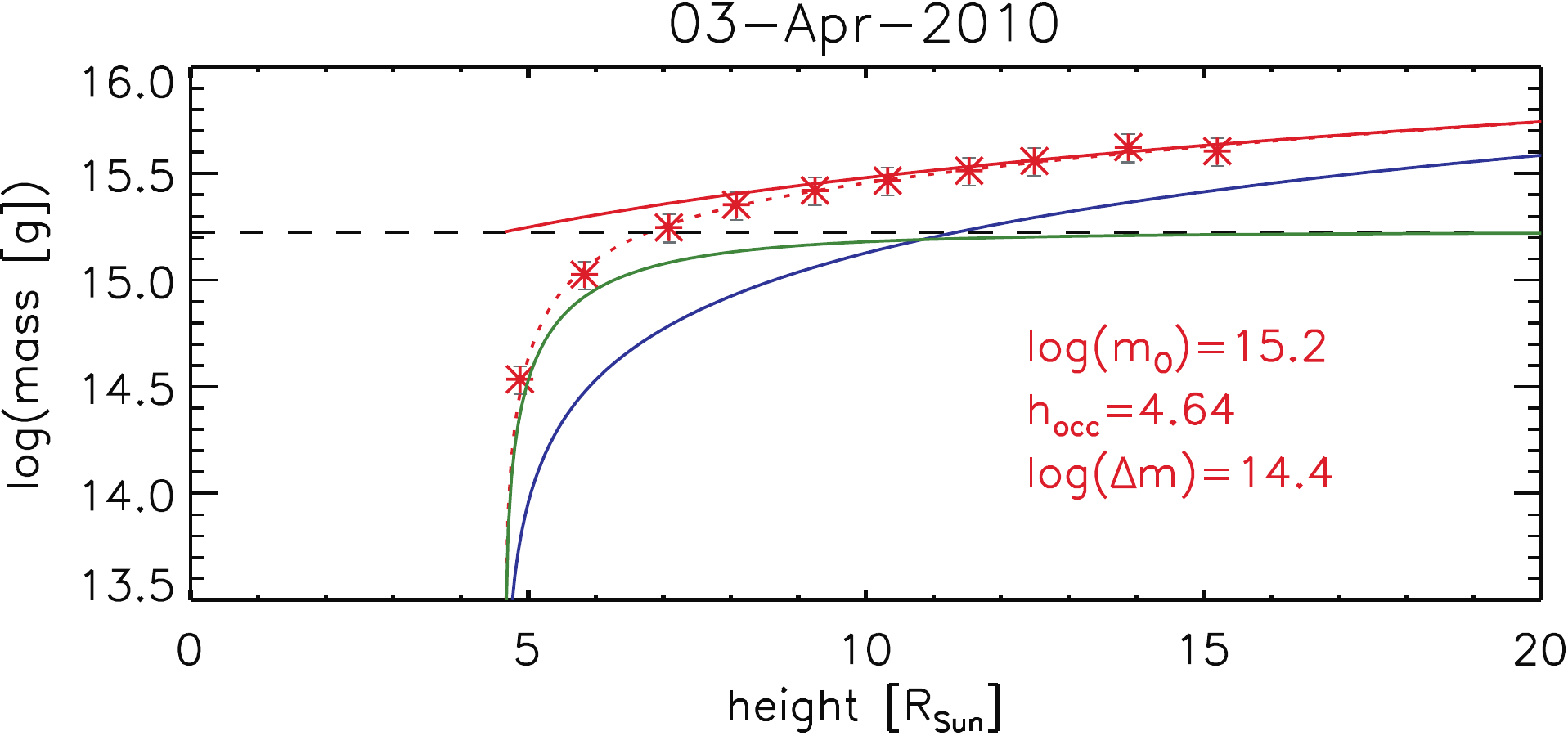}
	\caption{3D CME mass evolution for the 2010 April 3 event. The measurements (asterisks) together with the fit (dotted line) and the real mass evolution (solid line) are plotted in red. Additionally the two fit components are plotted separetely, the geometrical effect in green, the `real' mass increase as the blue solid line. The black dashed line represents the initial ejected mass $m_{0}$.}
	\label{fit0}
\end{figure}

Figure \ref{masstime} shows COR1 and COR2 mass measurements derived from STEREO-A (top panel) and STEREO-B observations (middle panel) and the 3D CME mass evolution (bottom) against time. COR1 and COR2 observations at the same time do not match because they do not include the same CME volume due to the different occulter sizes of the instruments. For example, a CME observed in COR2 below $\sim$5~R$_{\odot}$ shows less mass than observed in COR1 for the same height range. This occulter effect has been noted since the very first LASCO mass measurements \citep{vourlidas2000}. To test and properly quantify the occulter effect, we calculate the COR1 mass applying the COR2 occulter size and position (cf.\ Fig.\ \ref{occulter}). These measurements are plotted as green triangles in Fig.\ \ref{masstime} and fit well with the actual COR2 measurements. This exercise clearly illustrates the influence of the occulter size on CME mass measurements and suggests that those measurements, where the CME is only partially imaged, in any coronagraph, should be corrected via the procedure outlined below. \\
Figure \ref{massheight} shows the 3D CME mass evolution against deprojected height for the 2010 April 3 event. The observed mass increase at low heights is due to the gradual appearance of mass in the field-of-view (FoV) of the telescope. By comparing observations with theoretical considerations of this geometrical effect, we can derive a fit function, which describes our observations, and also enables us to estimate the mass hidden behind the occulter.
We describe the derivation of that fit function, which is based on the following assumptions:
\begin{itemize}
\item The volume of the CME has the shape of a spherical sector.
\item The CME expands self-similary and adiabatically.
\item The CME mass remains constant during the CME propagation from the low corona to the first coronagraph FOV.
\end{itemize}
We describe the volume of the CME with the formula of a spherical sector:
\begin{equation}
\begin{split}
V_{CME}=\int^{h}_0 \int^{\theta}_0 \int^{2\pi}_0 r^2 sin\theta~dr~d\theta~d\phi\\
=\frac{2\pi h^3}{3}(1-cos\theta),
\end{split}
\label{sector}
\end{equation}
where \textit{h} is the radius of the sphere (height of the CME leading edge measured from the center of the solar disk) and $\theta$ is the width of the spherical sector (CME width). Figure \ref{model1} shows an image of the 2010 April 3 event together with the boundary of a spherical sector outlined. Then the occulted CME volume, $V_{occ}$, is given by Eq.\ \ref{sector} for height $h=h_{occ}$, the effective occultation size:
\begin{equation}
V_{occ}=\frac{2\pi h_{occ}^3}{3}(1-cos\theta).
\end{equation}
Hence the observed CME volume is
\begin{equation}
V_{obs}(h)=V_{CME}-V_{occ}=\frac{2\pi}{3}(1-cos\theta)(h^3-h_{occ}^3).
\label{volume}
\end{equation}
Under the assumptions that i) the CME expands self-similarly, i.e.,\ the width $\theta$ remains constant during the evolution, and ii) the CME expands adiabatically, the density change can be expressed by
\begin{equation}
\rho (h)=\rho_0 \left(\frac{h_{occ}}{h}\right)^3
\label{density}
\end{equation}
with $\rho_0$ the density at the beginning of the CME evolution, calculated as $\frac{m_0}{V_{occ}}$. The parameter $m_0$ is the initially ejected mass, i.e.\ `true' CME mass at the time when the CME becomes the first time visible behind the occulting disk. Hence, the initially ejected mass $m_0$ can be calculated from the visible fraction of mass $m(h)$ which is not occulted. Using Eq.\ \ref{volume} and \ref{density} we derive the observed mass at a given height as
\begin{equation}
m(h)=\rho(h) \  V_{obs}(h)= m_0\left(1-\left(\frac{h_{occ}}{h}\right)^3\right).
\label{geo}
\end{equation}
Because the occultation height $h_{occ}$ is a constant, the second term of Eq.\ \ref{geo} becomes small with increasing CME height and $m(h)$ converges to $m_0$. At heights $h=h_{occ}$, $m(h)=0$, i.e.\ all the CME mass is hidden behind the occulter. At heights below $h_{occ}$ the function $m(h)$ of Eq.\ \ref{geo} is not defined.
\\
A closer look at the later phase of the CME mass evolution ($\sim$ 10 to 15 R$_{\odot}$), observable only in the COR2 FoV, shows a slight increase in mass, which cannot be explained by a pure geometrical effect due to the occulter (Eq.\ \ref{geo}). It could be due to a pile up, i.e.\ the CME sweeps up coronal material as it propagates through the corona. Another possibility could be a mass flow from the low corona in addition to $m_0$, when the CME is observable in the COR2 FoV. To take into account such effects, we add the term $\Delta m(h-h_{occ})$ to Eq.\ \ref{geo} to describe a real CME mass increase (in contrast to the geometrical effect). Including this term into Eq.\ \ref{geo} we obtain for the CME mass evolution with height
\begin{equation}
m(h)= m_0\left(1-\left(\frac{h_{occ}}{h}\right)^3\right)+\Delta m(h-h_{occ}),
\label{endformel}
\end{equation}
with $\Delta m$ the real mass increase per height. Whereas the term $m_0\left(\frac{h_{occ}}{h}\right)^3$ in Eq.\ \ref{endformel} becomes smaller with height, the last term increases. At the height $h_{occ}$, the observed mass is equal to zero, since all of the CME mass is occulted. We stress that Eq.~\ref{endformel} describes the \textit{observed} mass as a function of distance but with the knowledge of its fit parameters also the `true' mass evolution can be reconstructed. In Fig.\ \ref{fit0} the fit function applied to the COR2 observations of our example event is shown as the dotted red line. With this fit we find an initial ejected mass $m_0$ of $1.68\cdot10^{15}\pm2.25\cdot10^{14}$ g and a real mass increase of $2.50\cdot10^{14}\pm2.57\cdot10^{13}$ g/R$_{\odot}$. The two fit components (geometrical effect and actual mass increase) are plotted separetely. The red solid line shows the `true' CME mass evolution, corrected for the geometrical effect and calculated by
\begin{equation}
m(h)=m_{0}+\Delta m(h-h_{occ}).
\label{trueevolution}
\end{equation}
For a proper evaluation of the fit function we refer to the appendix of this paper.

\subsection{Mass evolution derived from fit function}

\begin{figure}
	\centering
		\includegraphics[scale=0.7]{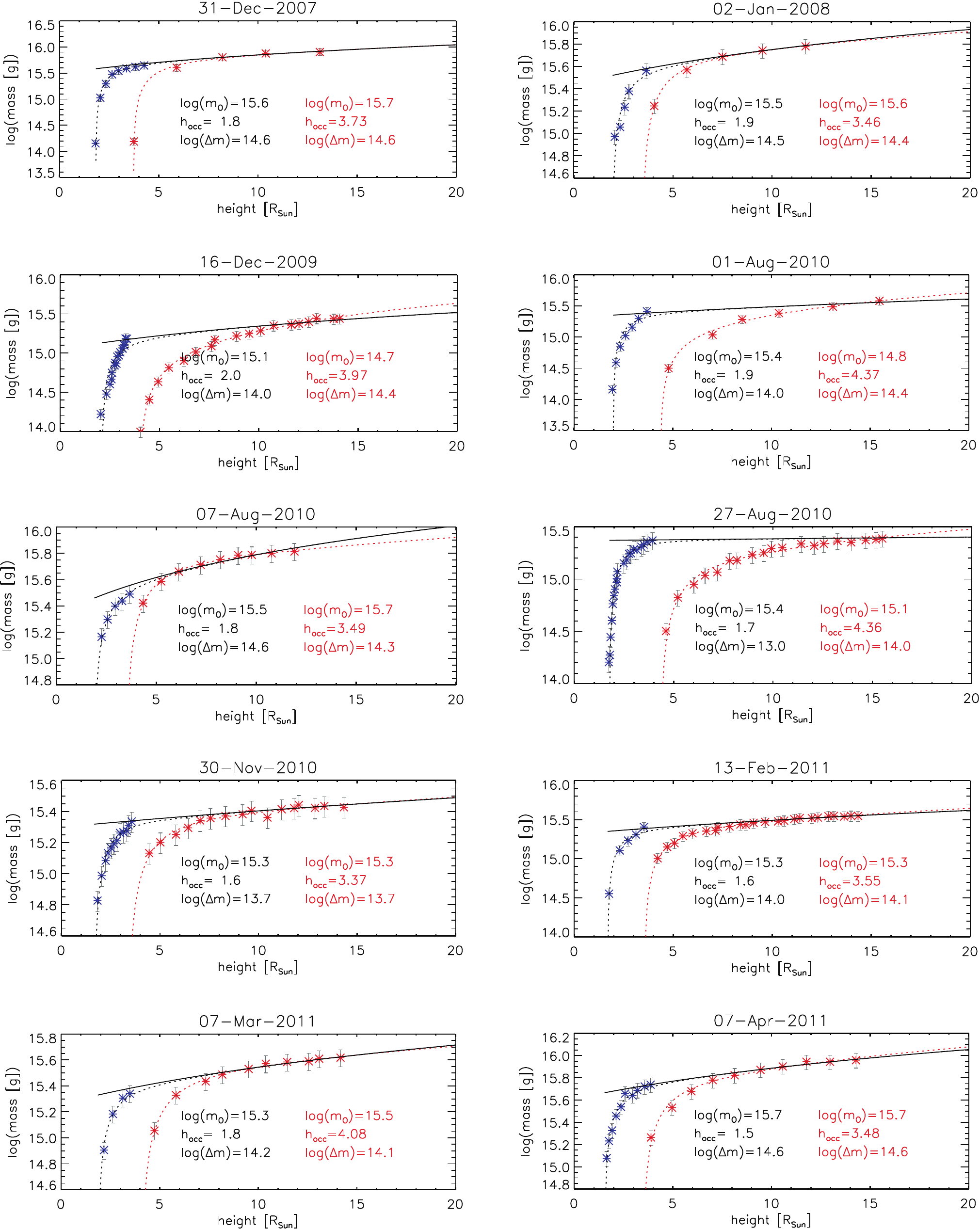}
	\caption{3D CME mass evolution against deprojected heights for 10 events out of our sample. Blue triangles represent COR1 observations, red asterisks COR2 measurements. The fit applied to COR2 observations is shown as the red dotted line together with its three fit parameters. The black dotted line represents the fit applied to combined COR1 and COR2 measurements (only COR2 data points were used, which are at least 10\% higher than the last COR1 data point). From the parameters of this fit (printed in black) the `true' CME mass evolution can be calculated (black straight line).}
	\label{fits}
\end{figure}

\begin{figure}
	\centering
		\includegraphics[scale=2.0]{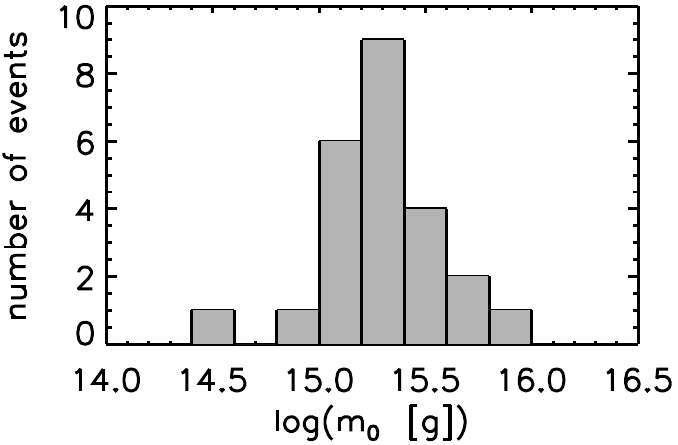}
	\caption{Ejected mass $m_{0}$ derived from the fit applied to combined COR1 and COR2 observations.}
	\label{histm0}
\end{figure}

\begin{figure}
	\centering
		\includegraphics[scale=2.]{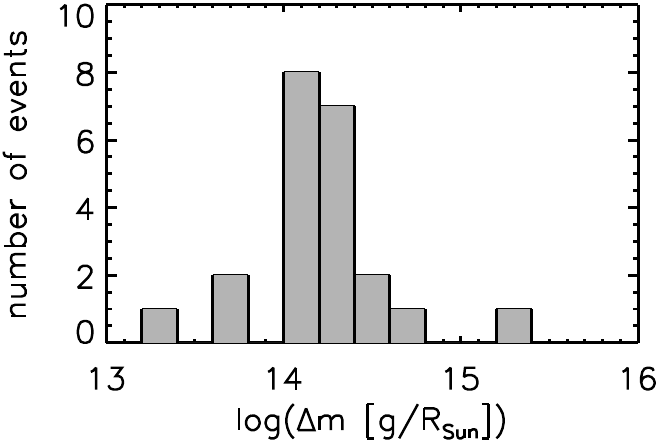}
	\caption{Distribution of the real CME mass increase rate $\Delta m$, derived from the fit applied to the COR2 mass measurements.}
	\label{histma}
\end{figure}

We derive the `true' mass evolution from a) combined COR1/COR2 measurements (applied on 24 out of 25 events - one event has too few data points) and b) solely from COR2 measurements (applied on 23 out of 25 events - two events have too few data points). Figure \ref{fits} shows the results for 10 events of our sample. We combine COR1/COR2 data, assuming that for COR2 the occulter effect becomes negligible at larger heights, by using all COR1 measurement points and only those COR2 measurements which are at least 10\% larger than the last measurement in COR1. The combined COR1/COR2 fit curves (Eq.\ \ref{trueevolution}) are represented in Fig.\ \ref{fits} as black dotted lines and demonstrate very well that it is possible to combine mass measurements derived from different instruments. With this we are able to calculate the `true' mass value from low coronal heights ($<3$~R$_{\odot}$) up to about 20~R$_{\odot}$.
\\
Comparing the fit parameter for the real mass increase, $\Delta m$, derived from COR1/COR2 and COR2 measurements (Eq.\ \ref{endformel}; shown as red dotted lines in Fig.\ \ref{fits}), we find for about 50\% of the studied events a good match with differences less than 20\%. This result is expected since we observe in both instruments the same CME. However, for some events there are large discrepancies. It turned out that only for well observed events, i.e.\ if coronal disturbances are small compared to the CME observation, the COR1/COR2 fit gives reliable results for the real mass increase. Therefore, we determine $\Delta m$ from the fit applied to COR2 observations and use the combined COR1/COR2 fit to get an estimate for the initial ejected mass $m_{0}$ for which we derive values ranging from $3.6\cdot10^{14}$ g to $8.9\cdot10^{15}$ g with a mean value of the logarithmic data of 15.32, corresponding to a mass value of $2.1\cdot10^{15}$ g. The distribution of all $m_{0}$ values is displayed in Fig.\ \ref{histm0}.
\\
Beside $\Delta m$, we derived from COR2 the fit parameters $m_{10}$, the mass value at 10 R$_{\odot}$, and $m_{end}$, the mass value at 20 R$_{\odot}$. For the calculation of $m_{10}$ and $m_{end}$ we use the `true' mass evolution (Eq.\ \ref{trueevolution}) after eliminating the geometrical effect. The statistical properties of fit parameters and CME masses are summarized in Table \ref{parametertable}. Since the CME mass values are lognormally distributed \citep[see][]{vourlidas2010}, the statistical quantities are calculated in a logarithmic space.
\\
Figure \ref{histma} shows the distribution of the real mass increase rate $\Delta m$ for our sample. Only one event out of 23 shows no mass increase in the late phase of the CME evolution. For $\Delta m$ we find a range of $2.4\cdot10^{13}$ to $2.4\cdot10^{15}$ g/R$_{\odot}$. The mean value of the logarithmic data is 14.25, corresponding to a mass increase of $1.8\cdot10^{14}$ g/R$_{\odot}$. The relative mass increase $\Delta m/m_{end}$ (see Fig.\ \ref{histmam10}) is on average 0.04, meaning that we have a real CME mass increase of about 4\% per R$_{\odot}$ up to 20 R$_{\odot}$. Minimum and maximum values are 0.02 and 0.06, respectively.

\subsection{Force and energy distributions}
\begin{figure}
	\centering
		\includegraphics[scale=2.0]{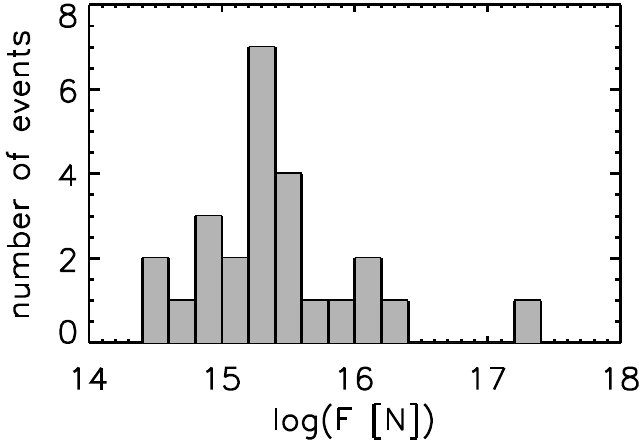}
	\caption{Distribution of total forces calculated by $m_{0} a_{max}$ for 24 events. $m_{0}$ is derived from the fit applied to combined COR1 and COR2 measurements, $a_{max}$ is the peak value of the CME acceleration profile.}
	\label{histforce}
\end{figure}

\begin{figure}
	\centering
		\includegraphics[scale=2.0]{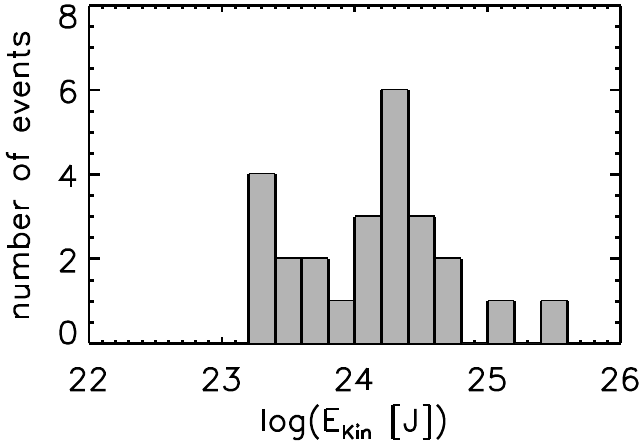}
	\caption{Distribution of the CME kinetic energy.}
	\label{histenergy}
\end{figure}

We use the CME peak accelerations $a_{max}$, peak velocities $v_{max}$ and the ejected masses $m_{0}$ to calculate total forces ($F=a_{max} m_{0}$) and kinetic energies ($E_{kin} = \frac{1}{2} m_{0} v_{max}^2 $) in the low corona. Since the distributions of the total forces and kinetic energies also follow a lognormal distribution, the distributions are plotted and their statistical properties are calculated in a logarithmic scale. The statistical properties are summarized in Table \ref{parametertable}.
\\
There are different forces (Lorentz force, effective drag force, gravitational force), which act on the CME and causing its acceleration or deceleration \citep{vrsnak2006}. In this work, we don't distinguish among the different forces but observe the effect of the net force. Figure \ref{histforce} shows the distribution of the maximal total forces. We note that $a_{max}$ values are measured between 1.1 and 3.3 R$_{\odot}$, whereas $m_{0}$ is assumed as the mass the CME has reached in the height range between 1.4 and 2.8 R$_{\odot}$ (derived from the fit). But because the peak acceleration for many CMEs occurs at low heights \citep[see][]{bein2011} that they are inaccessible by coronagraphs, $m_{0}$ is the best available estimate for the CME masses at that heights. This is confirmed by the \citet{aschwanden2009} study, who did not find large differences in the masses, derived from EUVI and COR1 observations, respectively. The distribution in Fig.\ \ref{histforce} shows that the derived net force ranges from values of $2.2\cdot10^{14}$ to $5.1\cdot10^{16}$ N. We find for the logarithms of the total forces a mean value of 15.18, corresponding to $1.5\cdot10^{15}$ N.
\\
We also use $m_{0}$ to calculate the kinetic energy $E_{kin}$. $v_{max}$ is measured at heights between 1.5 and 8.4 R$_{\odot}$, but the velocity does actually not change that much during the evolution in the coronagraphic FoVs. In Fig.\ \ref{histenergy} the distribution of the kinetic energies is plotted, which range from $1.0\cdot10^{23}$ to $6.6\cdot10^{24}$ J. We find a mean value of 23.89 derived from the logarithmic data, corresponding to $7.8\cdot10^{23}$ J.

\section{Summary and Discussion}

\begin{figure}
	\centering
		\includegraphics[scale=0.75]{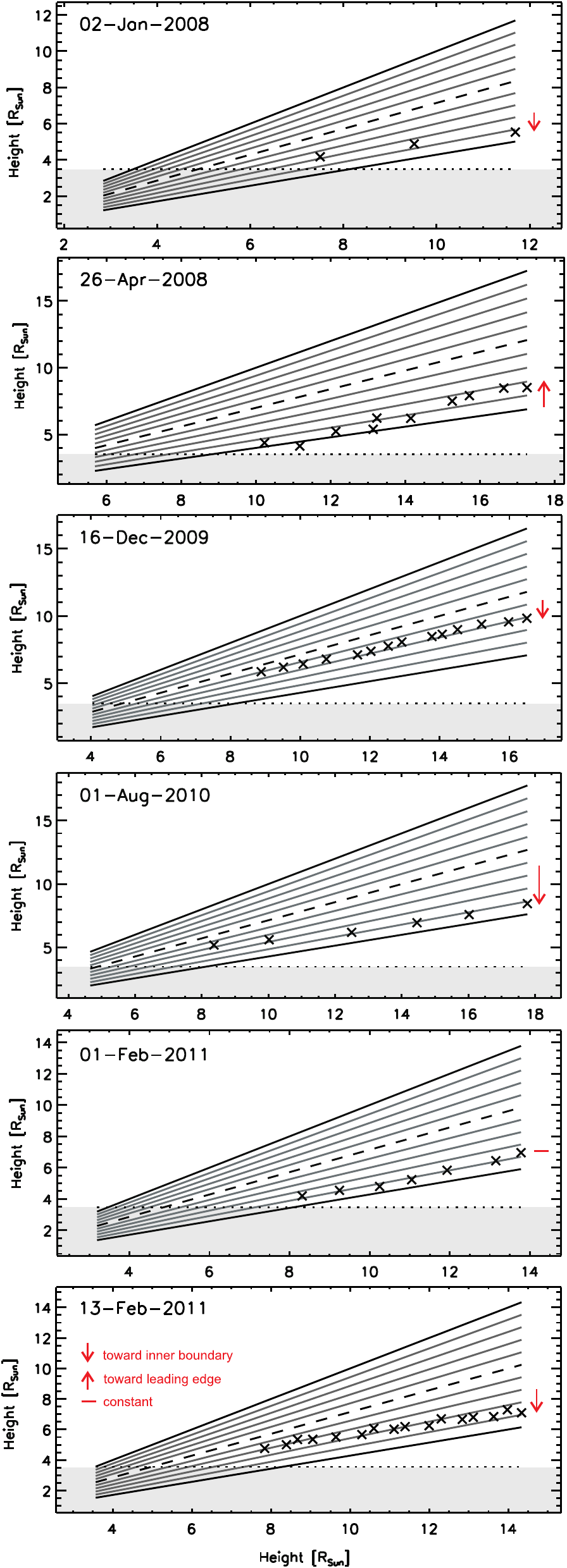}
	\caption{Center of mass evolution within the CME flux rope for example events. The center of mass evolution against the leading edge height is represented by crosses, the outer and inner boundary of the CME flux rope are plotted as black solid lines. The geometric flux rope center (dashed line) and grey lines in 10\% steps of the flux rope width are plotted to better track the center of mass motion (crosses) within the flux rope. The dotted line represents the effective occultation height $h_{\rm occ}$ derived from the fit.}
	\label{fl}
\end{figure}

In this paper we presented the 3D CME mass evolution against time and deprojected heights for a sample of 25 events, which were observed by the COR1 and the COR2 instrument onboard both STEREO s/c.
\\
The CME mass evolutions show a strong increase at the beginning of the evolution, which was also found in former studies \citep{vourlidas2000, colaninno2009, carley2012} and is explained by a geometrical effect, i.e.\ by a gradual appearance of CME material above the occulting disk in the course of CME expansion. This occulter effect is confirmed by finding a match between temporally overlapping COR1 and COR2 mass measurements when applying the same occulter size for both instruments (cf.\ Fig.~\ref{masstime}).
\\
However, the observed mass increase in the later phase of the CME evolution ($\sim$ 10--15 R$_{\odot}$) is not due to mass hidden behind the occulter but a real increase, which can be explained either by pile up at the CME leading edge or continuous mass flow from the lower atmosphere. We developed a fit function taking into account both effects of mass increase (geometrical and physical) and found a very good match between fitted and observed data points. This function returns three fit parameters: (1) the effective occultation size of the instrument (depending on the direction and the CME expansion), (2) the initially ejected mass of the CME and, (3) the real mass increase per height. In the appendix we show all three are physically meaningful parameters. With Eq.\ \ref{trueevolution} we can estimate the `true' mass evolution and thus can derive the `true' mass value at every height larger than the effective occultation size. By this we can calculate $m_{end}$, defined as the mass at 20~R$_{\odot}$ and find that it is the most appropriate measure for the mass in the late phase of the CME evolution for three reasons: i) $m_{end}$ is always related to the same height, ii) $m_{end}$ is derived from the fit, which considers the shape of the CME mass-height curve (e.g.\ a probable mass increase), iii) at $\sim$20~R$_{\odot}$ the CME reaches the Alfv$\acute{\rm e}$nic critical point beyond which it can be assumed that the evolution more or less ceases \citep{hundhausen1972}.
\\
We have shown that mass measurements derived from different instruments can be combined even if the coronagraphs have different occulter sizes. We derived from combined COR1 and COR2 observations initially ejected masses $m_{0}$ in the range of $3.6-89\cdot10^{14}$ g. This fit parameter gives us an estimate of CME mass low in the corona ($<$3~R$_{\odot}$) which may be important for calculating initial forces acting on the CME, since the CME peak accelerations occur also at similar heights low in the corona \citep{bein2011}. For the total forces we found values between $2.2\cdot10^{14}$ and $5.1\cdot10^{16}$ N. Kinetic energy values were found to lie between $1.0\cdot10^{23}$ and $6.6\cdot10^{24}$ J. We note here that these deprojected quantities are in good agreement with the overall CME statistics \citep{vourlidas2010} which are based on projected quantities (as do most of the relevant work before 2007). This agreement gives us confidence that  large statistical studies provide robust information for the actual physical parameters of CMEs even though they are based on single viewpoint measurements. \citet{emslie2004} showed that CME kinetic energy values can be even higher. They presented projected kinetic energy values of $1.8\cdot10^{25}$ J and $1.1\cdot10^{25}$~J for two strong CME/flare events. \citet{carley2012} presented a case study concerning the deprojected energy and force evolution of the 2008 December 12 CME event and found values of $6.3\pm3.7 \cdot10^{22}$ J for the energy and $3.9\pm5.4 \cdot10^{13}$ N for the force at about 3 R$_{\odot}$, which is small compared to our study. That can be explained by the fact that our sample includes mostly events during the ascending phase of the solar cycle and, thus, have on average larger velocities, accelerations and masses.
\\
The derived mass increases per R$_{\odot}$ were found to be between 2\% and 6\% up to 20 R$_{\odot}$, corresponding to $2.4\cdot10^{13}$ g/R$_{\odot}$ to $2.4\cdot10^{15}$ g/R$_{\odot}$. It is well known that absolute mass measurements are subject to errors of up to a factor of two \citep{vourlidas2000, lugaz2005, vourlidas2010}. However, here we were mostly interested in the relative change of CME masses, not in the absolute calibration, and the relative error is much smaller. If the derived mass increases were only due to errors, then we would expect both increases and decreases of mass across different events. This is not the case. \citet{lugaz2005} reported from 3D MHD simulations of CMEs a mass increase of the order of 13\% within 1 to 2.5 hours after the CME initiation and a weaker one up to 1 AU. The authors also compared the results from 3D simulations with mass measurements from line-of-sight images and found for the latter a weaker mass increase, which they explained by the following: When measuring the mass from a single viewpoint coronagraphic image, it is assumed that all mass is placed on the POS, which is not true for the whole volume of the CME, leading to an underestimation of the mass. Because of the expansion of the CME, this effect becomes larger with height. More electrons deviate from the POS with larger distances making it difficult to distinguish a mass increase against the reduced brightness. Thus, the mass increases found in our study can be considered as a lower limit.
\\
These real CME mass increases can be explained by a pile up of mass during the propagation/expansion of the CME through the corona, as suggested by simulations from \citet{das2011}, who showed that a pile up is possible also at lower heights (2 to 7 R$_{\odot}$). However, no such effect has been observed so far in the 2.2 - 30~R$_{\odot}$ range \citep{howard2005}.
On the other hand, mass could be supplied to the CME from mass flow resulting from the opening of the field and the creation of a temporary coronal hole behind the CME in the low solar corona. Coronal dimming regions analysed by former studies showed mass outflows lasting for hours \citep{reinard2008, aschwanden2009, bemporad2010, miklenic2011, tian2012} and thus it is likely that they provide the CME with material also in its later evolution. \citet{vourlidas2000} already pointed out this effect to explain the backward (relative to the CME front) motion of the CME center of mass as the event expands in the outer corona.
\\
The behavior of the center of mass of the CME during its evolution should clarify which of the processes described above is responsible for the real CME mass increase observed. In case of pile up, more CME material accumulates at the front of the CME and the center of mass is expected to move toward the CME leading edge with time. If there is a mass flow from behind, the center of mass would drop backward with time. From the COR2 observations of our events we calculated for each time step the center of mass position within the defined CME sector following \citet{vourlidas2000}. The distance of the center of mass to the Sun center is then plotted together with the height of the CME leading edge, which we consider as the outer boundary of the modeled CME flux rope (cf.\ Fig.\ \ref{fl}). The inner boundary of the flux rope was calculated by the width derived from the forward modeling. The distance of both boundaries (black solid lines in Fig.\ \ref{fl}) increases because of the self-similar expansion of the flux rope. Because a significant fraction of the flux rope is occulted in the beginning of the CME evolution, the center of mass was not measured until the full flux rope could be observed, i.e.\ the height of its inner boundary is larger than the effective occultation size (indicated by the shaded area below the dotted line in Fig.\ \ref{fl}).
Fig.\ \ref{fl} indicates with grey lines steps of 10\% of the width of the CME body. From this we obtain that on average, the center of mass lies on the 25\% line and shifts with 5$\pm$7\% towards the inner boundary of the flux rope. For 14 events the center of mass moved with more than $5\%$ of the flux rope width toward the inner boundary (mass flow), in 2 cases we measured a motion over 5--10\% of the CME width toward the outer boundary (pile up). For 7 cases the center of mass motion relative to the CME leading edge remains roughly constant (i.e.\ changes are $<5\%$ of the flux rope width) and for 2 events we cannot make a clear statement. Therefore, we find that mass flow behind the CME is the main reason for the mass increase.
\\
Prominences, which are often related to CMEs, can also influence the CME mass measurements. In some cases their bright H$\alpha$ radiation can be observed in white light coronagraphic observations. When this radiation is wrongly interpreted as Thomson scattering, excessive mass values are obtained. \citet{carley2012} explained the short-time increase and decrease of mass in the COR1 measurements of their event under study by this phenomenon. Therefore we took special care to exclude events which showed H$\alpha$ radiation in the coronagraphic observations.
Nevertheless H$\alpha$ emission could be one reason for the differences in the mass increases derived from the COR2 fit and the combined COR1/COR2 fit (see for example 2010 August 27 in Fig.\ \ref{fits}). Scattering effects of the COR1 instrument or mass draining down before it reaches the COR2 FoV can also lead to larger mass measurements in the COR1 FoV. Coronal disturbances, e.g.\ other CMEs, background streamers or particles produced by a solar flare and hitting the detector may influence COR1 measurements as well as COR2 measurements.

\section{Conclusions}

The observed CME mass evolution in coronagraphic observations is easily explained by the `true' mass evolution and a geometrical effect. We derived a fit function, which enables us to consider both components separately. By this `true' mass values at every height, between the effective occultation size and 20~R$_{\odot}$, can be estimated. For the majority of the events we find a significant CME mass increase, which is in the range of 2 to 6\% and becomes most important over the distance range 10--20~R$_{\odot}$. We find that most of the mass is located in the rear part of the CME body and, in about half of the events, there is strong evidence that the derived mass increase is supplied to the back of the CME by enhanced flow from the low corona due to temporary coronal holes created by the CME eruption. Pile-up of coronal material ahead the CME is rare in agreement with the findings of \citet{howard2005} results.

\acknowledgements
This work is supported by the \"Osterreichische F\"orderungsgesellschaft (FFG) of the Austrian Space Applications Programme (ASAP) under grant no. 828271 and by the Fonds zur F\"orderung wissenschaftlicher Forschung (FWF): V195-N16. The work of Angelos Vourlidas is supported by NASA contract S-136361-Y to the Naval Research Laboratory. We want to thank Robin Colaninno for providing the mass calculation programs.
The STEREO/SECCHI data are produced by an international consortium of the Naval Research Laboratory (USA), Lockheed Martin Solar and Astrophysics Lab (USA), NASA Goddard Space Flight Center (USA), Rutherford Appleton Laboratory (UK), University of Birmingham (UK), Max-Planck-Institut f\"ur Sonnensystemforschung (Germany), Centre Spatiale de Li\`{e}ge (Belgium), Institut d'Optique Th\'{e}orique et Appliqu\'{e}e (France), and Institut d'Astrophysique Spatiale (France).

\begin{table*}
	\centering
	\small
		\begin{tabular}{| c| c | c | c| c |c |}
		\hline
		 &&&&\\
			 & mimimum & maximum & arithmetic mean $\pm$   & median $\pm$ median \\
			 &  &  &    standard deviation &  absolute deviation    \\
			&&&&\\
			\hline
			&&&&\\
			$log(m_{0} [g])$ &  14.56 & 15.95 & 15.32 $\pm$ 0.29 & 15.35 $\pm$ 0.18 \\
			$h_{\rm occ} [R_{\odot}]$ &  2.69 & 5.08 & 3.92 $\pm$ 0.61 & 3.82 $\pm$ 0.36 \\
			$log(\Delta m [g/R_{\odot}] )$  &  13.38 & 15.39 & 14.25 $\pm$ 0.38 & 14.24 $\pm$ 0.14 \\
			$log(m_{10} [g])$ &  14.89 & 16.45 & 15.53 $\pm$ 0.34 & 15.53 $\pm$ 0.23 \\
			$log(m_{end} [g])$ &  15.11 & 16.72 & 15.72 $\pm$ 0.33 & 15.70 $\pm$ 0.20 \\
			$log(F [N]$) &  14.34 & 16.71 & 15.18 $\pm$ 0.56 & 15.27 $\pm$ 0.41 \\
			$log(E_{\rm kin} [J])$ &  23.00 & 24.82 & 23.89 $\pm$ 0.48 & 24.00 $\pm$ 0.42\\
			&&&&\\
		 \hline
	
		\end{tabular}
		\caption{Statistical quantities derived for different CME mass and fit parameters. $m_{0}$ is a parameter from the fit applied to combined COR1 and COR2 observations. $h_{\rm occ}$ and $\Delta m$ are fit parameters of the fit applied to COR2 observations. $m_{10}$ and $m_{end}$ are derived from the same fit, representing the CME mass at 10 and 20 R$_{\odot}$, respectively. The force is calculated by $F=m_{0} a_{max}$, the kinetic energy is calculated by $E_{\rm kin}=\frac{m_{0} \ v_{max}^2}{2}$.}
		\label{parametertable}
\end{table*}

\clearpage

\clearpage

\clearpage

\clearpage

\clearpage

\clearpage

\clearpage

\clearpage

\clearpage

\clearpage

\begin{figure}
	\centering
		\includegraphics[scale=2.0]{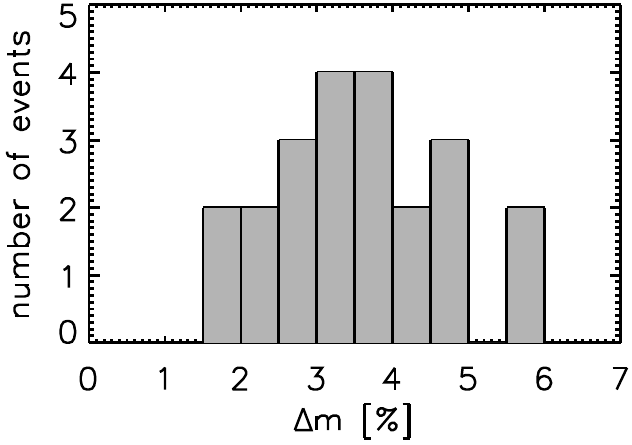}
	\caption{Distribution of the relative mass increase per R$_{\odot}$ defined as $\Delta m/m_{end}*100$.}
	\label{histmam10}
\end{figure}
\clearpage

\clearpage

\clearpage

\clearpage

\appendix

\section{Evaluation of the fit}

\begin{figure}
	\centering
		\includegraphics[scale=2.0]{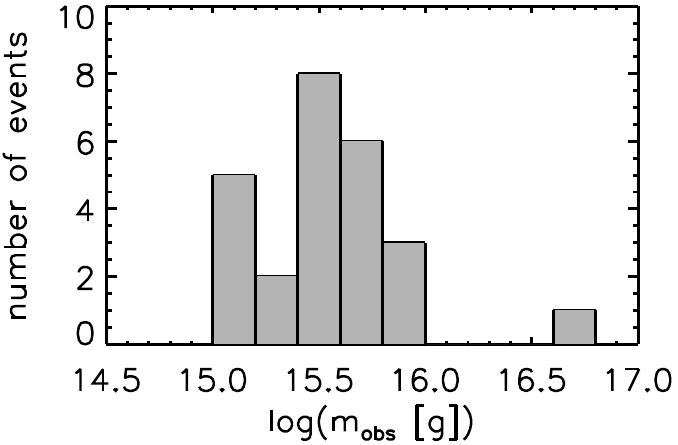}
	\caption{Distribution of $m_{obs}$, the mean of the last three CME mass measurements derived from COR2 observations for a sample of 25 events.}
	\label{histmm}
\end{figure}

\begin{figure}
	\centering
		\includegraphics[scale=1.5]{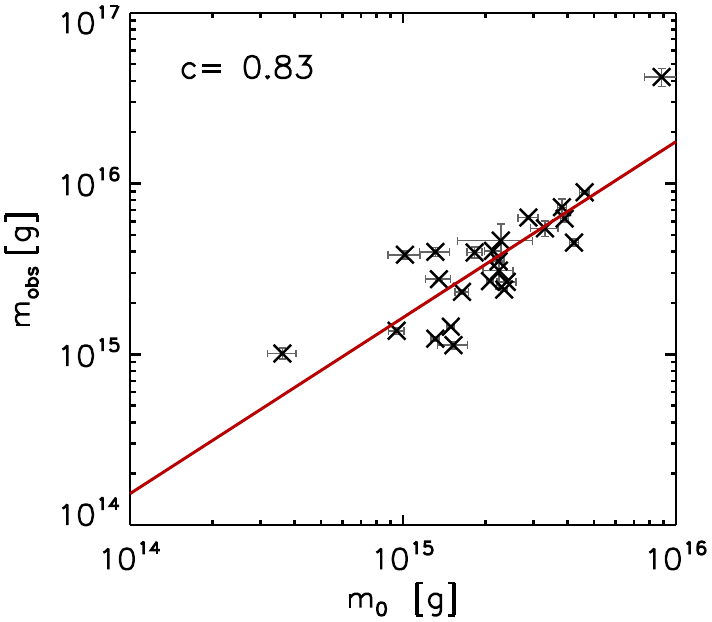}
	\caption{Ejected CME mass $m_{0}$ against $m_{obs}$, the mean value of the last 3 mass measurements of COR2 observations. The regression line is plotted in red. $m_{0}$ is derived from the fit applied to combined COR1 and COR2 measurements.}
	\label{m0mm}
\end{figure}

\begin{figure}
	\centering
		\includegraphics[scale=1.5]{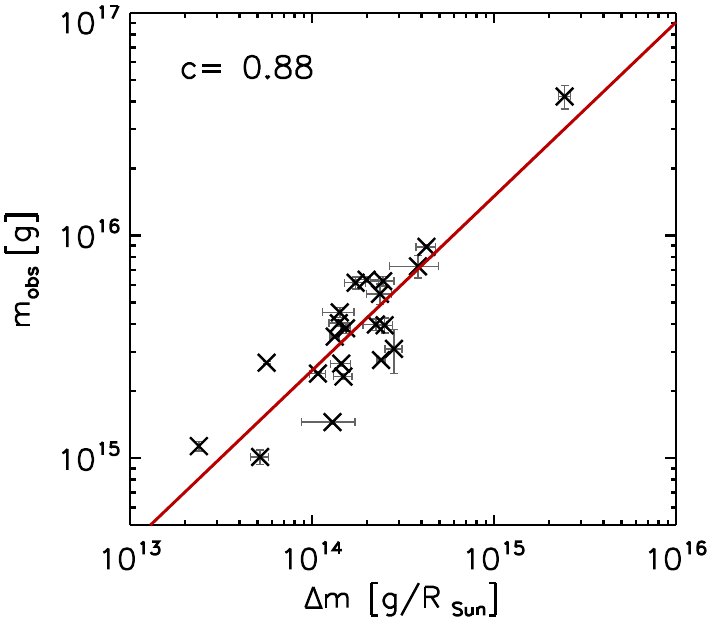}
	\caption{Real mass increase rate $\Delta m$ (derived from the fit applied to the COR2 mass measurements) against $m_{obs}$, the mean value of the last three COR2 measurements. The regression line (red solid line) and the correlation coefficient $c$ are calculated in double logarithmic space.}
	\label{mmma}
\end{figure}

\begin{figure}
	\centering
		\includegraphics[scale=0.7]{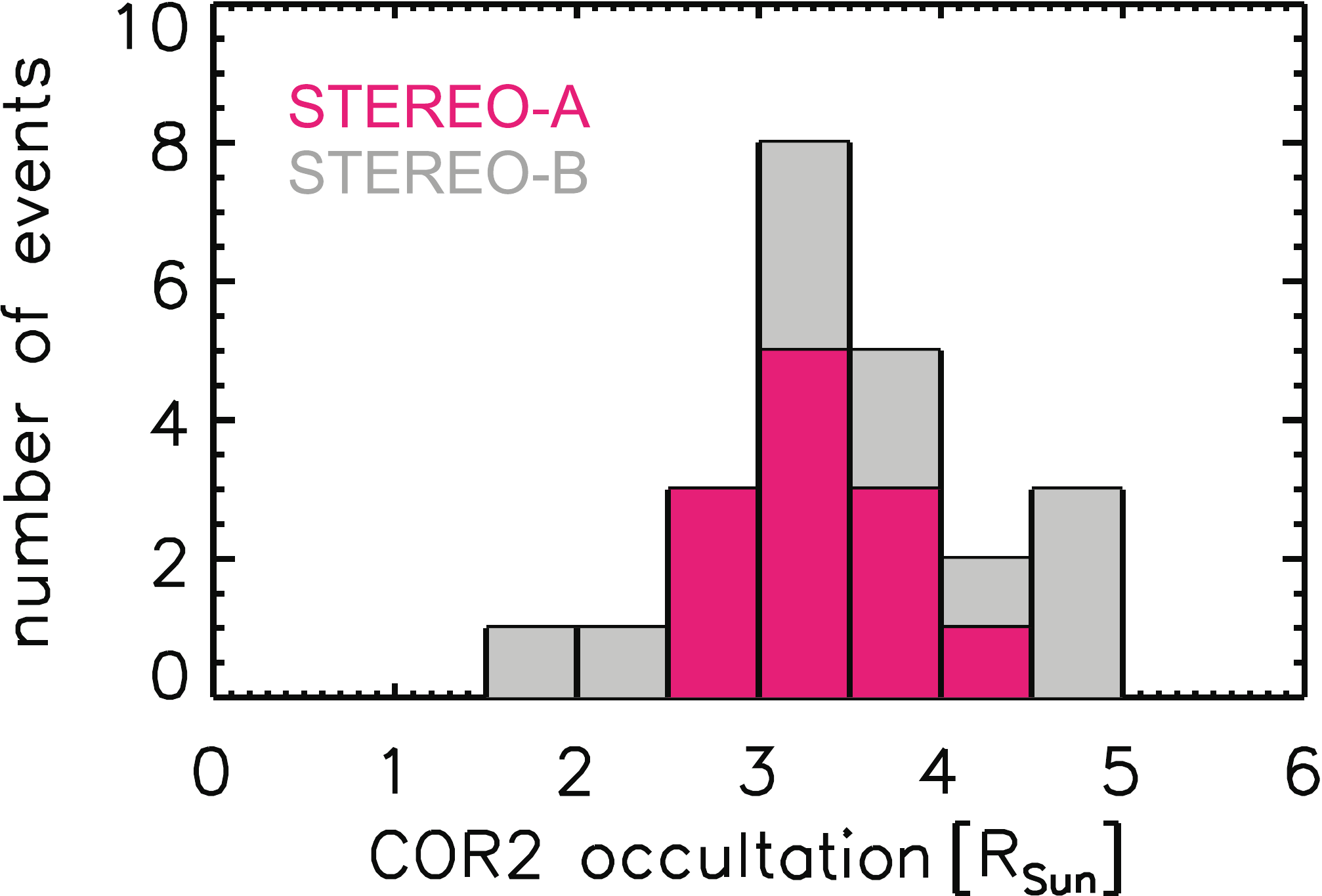}
	\caption{The effective occultation size $h_{\rm occ}$ is derived from the fit, which is applied to 3D COR2 CME masses against deprojected heights. The distribution shows the projection of the $h_{\rm occ}$ values on the POS corresponding to the CME propagation direction. STEREO-A occultation sizes are plotted in color, STEREO-B occultation sizes in grey.}
	\label{occulter2}
\end{figure}

In the following we evaluate our fit function against the observational parameters. The fit parameters are the effective occultation size of the instrument $h_{ \rm occ}$, the initially ejected mass of the CME $m_{0}$ and the real mass increase per height $\Delta m$. As observational parameters we use the actual size of the occulter disk and $m_{\rm obs}$.
\\
Due to the evolving CME mass with height, we define m$_{obs}$ as the mean value of the last three COR2 measurements, which are measured in the outer range of the COR2 FoV. Figure \ref{histmm} shows the distribution of all derived m$_{obs}$ values, plotted on a logarithmic scale. From the logarithmic data, $log(m_{obs})$, we find a mean value of 15.55, which corresponds to a mass of $3.6\cdot10^{15}$~g. Minimum and maximum values of m$_{obs}$ measurements are $1.0\cdot10^{15}$ g and $4.2\cdot10^{16}$ g, respectively.
\\
In the course of the derivation of the fit function we assume that $m_{0}$ is the initally ejected mass, i.e.\ a physical meaningful measure. To confirm our assumption we compare $m_{0}$ with the measured mass values $m_{obs}$ (see Fig.\ \ref{m0mm}). Because $m_{0}$ represents the mass at about 1.5 to 3~R$_{\odot}$ and $m_{obs}$ the mass at about 15~R$_{\odot}$ these two measurements should be highly correlated since we observe a linear real mass increase $\Delta m$. Indeed we find a positive correlation with a correlation coefficient of 0.83.
\\
We also find a positive correlation between $m_{\rm obs}$ and the real mass increase $\Delta m$ of $c=0.88$ shown in Fig.\ \ref{mmma}, which means that more massive CMEs also have a larger real mass increase. On the other hand, the relative mass increase $\Delta m/m_{end}$ (see Fig.\ \ref{histmam10}) shows no correlation with $m_{\rm obs}$.
\\
We derive the effective occultation size $h_{ \rm occ}$ from the fit applied to deprojected data, thus $h_{ \rm occ}$ is larger than the physical radius of the occulter dependent on the deviation of the CME propagation direction from the POS. Considering the CME propagation direction, $h_{ \rm occ}$ gives us meaningful values for the effective size of occultation. Fig.\ \ref{occulter2} shows the distribution of the occultation size projected back on the spacecraft POS.

\clearpage

\clearpage

\clearpage

\clearpage

\bibliographystyle{aa}

\end{document}